\documentclass[amsmath,amssymb,reprint]{revtex4-1}
\usepackage{graphicx}
\usepackage{dcolumn}
\usepackage{bm}
\newcommand{\p}{$\%$}

\newcommand{\gt}{$\gamma^{\prime\prime\prime}$}
\newcommand{\gd}{$\gamma^{\prime\prime}$}

\newcommand{\gmm}{$\alpha^{\prime\prime}\mathrm{-Fe_{16}N_{2}}$}

\newcommand{\eFeyN}{$\varepsilon-\mathrm{Fe_{3-y}N}$}
\newcommand{\jFeN}{$\zeta-\mathrm{Fe_{2}N}$}

\newcommand{\tfn}{$\mathrm{Fe_{4}N}$}

\newcommand{\muB}{$\mathrm{\mu_{B}}$}

\newcommand{\Ts}{$\mathrm{T_{s}}$}
\newcommand{\Tn}{$\mathrm{T_{N}}$}
\newcommand{\FeNx}{FeN$_\mathrm{x}$}
\newcommand{\mm}{$\mathrm{\textbf{M}}$}
\newcommand{\iFe}{$^{57}$Fe}

\begin{document}

\title{Magnetism and structure of \textit{in-situ} grown FeN films studied using N K-edge XAS and nuclear resonance scattering }

\author{Mukul Gupta} \email{mgupta@csr.res.in} \author{Nidhi Pandey, Niti, V. R. Reddy and D. M. Phase}
\address{UGC-DAE Consortium for
Scientific Research, University Campus, Khandwa Road, Indore 452
001, India}

\author{Kai Schlage, Hans-Christian Wille}

\address{Deutsches Elektronen-Synchrotron DESY, Notkestrasse 85,
D-22607 Hamburg, Germany}

\author{Ajay Gupta}

\address{Amity Center for Spintronic Materials, Amity University,
Sector 125, NOIDA 201 303, India}

\date{\today}

\begin{abstract}
We studied the structural and magnetic properties of
\textit{in-situ} grown iron mononitride (FeN) thin films. Initial
stages of film growth were trapped utilizing synchrotron based
soft x-ray absorption near edge spectroscopy (XANES) at the N
$K$-edge and nuclear resonant scattering (NRS). Films were grown
using dc-magnetron sputtering, separately at the experimental
stations of SXAS beamline (BL01, Indus 2) and NRS beamline (P01,
Petra III). It was found that the initial stages of film growth
differs from the bulk of it. Ultrathin FeN films, exhibited larger
energy separation between the t$_{2g}$ and e$_g$ features and an
intense e$_g$ feature in the N $K$-edge pattern. This indicates
that a structural transition is taking place from the rock-slat
(RS)-type FeN to zinc-blende(ZB)-type FeN when the thickness of
films increases beyond 5\,nm. The behavior of such N $K$-edge
features correlates very well with the emergence of a magnetic
component appearing in the NRS pattern at 100\,K in ultrathin FeN
films. Combining the \textit{in-situ} XANES and NRS measurements,
it appears that initial FeN layers grow in RS-type structure
having a magnetic ground state. Subsequently, the structure
changes to ZB-type which is known to be non-magnetic. Observed
results help in resolving the long standing debate about the
structure and the magnetic ground state of FeN.
\end{abstract}
\maketitle

\section{Introduction}
\label{intro}

Iron and nitrogen are among the most abundant elements on the
earth and therefore iron-nitrogen compounds - \FeNx, are not only
economical but also also environment friendly. They are being
investigated for applications in a variety of fields. For x $\leq$
0.5 in \FeNx, compounds formed are magnetic and have been explored
for applications as rare earth free permanent magnets, spintronic
materials and also other emerging magnetic
devices~\cite{Bhattacharyya_FeN_Review}. At x = 0.12 (in \FeNx),
the phase formed is \gmm, which has been long debated to possess a
giant magnetic moment, \mm$\sim$3\muB/Fe
atom~\cite{2000_JMMM_Fe16N2_gmm_or_not} and is still being a
subject of intense research~\cite{2016_Sci_Rep_Fe16N2}. At x =
0.25 the phase formed is \tfn~which is predicted to have almost
100\p~spin polarization ratio and a half-metallic
character~\cite{2006_PRB_Fe4N_SPR}, its larger (than Fe)
\mm~$\sim$2.5\muB/Fe
atom~\cite{JMMM10_Matar,APL11_Ito_Fe4N,2009_PSS_B_Fe4N_th_exp_Blanca}
makes it a suitable candidate in spintronic devices and a
potential alternative to Heusler
alloys~\cite{2018_APL_Fe4N_underlayer}. Between x = 0.33 to 0.5 in
\FeNx, phases formed are \eFeyN~(0 $<$ y $<$ 1), they are
thermally stable and possess soft-magnetic
properties~\cite{2016_JAC_Fe3N_soft_mag} and recently exhibited
for usage in Li-ion
batteries~\cite{2017_nano_energy_Fe3N_li-on_battery}. At x = 0.5,
the \jFeN~phase forms~\cite{Cai_2000} which has been used as a
precursor to produce iron mononitride (FeN) and iron pernitride
(FeN$_2$) by application of high-pressure and high-temperature
(HPHT)~\cite{2018_inorgchem_FeN2}.

N rich phases of iron-nitrogen compounds (x $>1$ in \FeNx) were
theoretically predicted for a while, e.g. the spinel nitride
Fe$_3$N$_4$~\cite{Fe3N4:JAP02,2002_APL_Fe3N4} and the pernitride
FeN$_2$~\cite{2011_Chemistry_FeN2_prediction}, only very recently
the FeN$_2$ phase has been synthesized experimentally under HPHT
(T$\sim$2000\,K,
P$\sim$50\,GPa)~\cite{2018_inorgchem_FeN2,2018_Nat_Com_poly_FeNx}
and by raising the pressure above 100\,GPa, the FeN$_4$ phase was
also evidenced by Bykov et al.~\cite{2018_Nat_Com_poly_FeNx}. Very
recently even higher N phases like FeN$_6$ and FeN$_8$ have been
predicted ~\cite{2018_Chemistry_Materials_polyN_FeN_theory}.
Polymeric nitrogen chains present in N rich \FeNx~compounds makes
them very attractive as high energy density materials (HEDMs),
however their synthesis under ambient temperature and pressure
remains a challenge.

While magnetic \FeNx~compounds already started to get attention in
early 1950s~\cite{Jack1951}, experimental works on the mononitride
phase can only be traced back in last three decades or
so~\cite{Oueldennaoua.SM.1985,Nakagawa.HypInt.1991,Suzuki.JAC1993,1997_PhysicaB_Hinomura}.
FeN compounds were extensively studied by Schaff et
al.~\cite{SchaafHypInt95,Niederdrenk_JAC96,Rissanen_JAC98,Schaaf.PMS.2002}
in late 1990s. Subsequently, FeN thin films were synthesized using
ion beam sputtering~\cite{2001_Vacuum_MG_FeN_IBS}, dc/rf magnetron
sputtering~\cite{2003_ASS_FeN_dcMS,2004_JAC_FeN_plasma_gas_cond,Gupta:PRB05,2006_TSF_FeN_TEM_Demange,Jouanny2010TSF,gupta:JAP2011,MG:JAC:2011,2015:PRB:MG:FeN,2015_JAC_FeAlN_MG,2017_Mat_today_FeN_dcMS,2019_JMMM_Fe4N_by_FeN},
pulsed laser deposition
(PLD)~\cite{Gupta_JAC01,2005_MSE_B_Laser_Nitriding,2012_HI_FeN_PLD,2013_HI_FeN_PLD},
high power impulse magnetron sputtering~\cite{SCT15_Akhil},
nitrogen plasma assisted molecular beam epitaxy
(MBE)~\cite{Navio.PRB08,2003_HI_Borsa,2006_NIMB_stoichiometic_FeN,JPak_2010,JCG15_FeN_MBE_Naito}
and very recently under
HPHT~\cite{2018_inorgchem_FeN2,2018_Nat_Com_poly_FeNx,2017_Iorg_Chem_FeN_FeN2,2018_JAC_FeN_megabar,2017_Angew_Chem_NiAs_FeN}.
From applications points of view, the mononitride FeN is also very
interesting as its oxidation resistance makes it a effective
catalyst in chemical
reactions~\cite{2006_Cat_Comm_FeN_Catalysis,2011_EES_FeN_Catalyst},
it can be used as precursor to yield magnetic phases in a
controlled
way~\cite{2019_JMMM_Fe4N_by_FeN,Navio.PRB08,Navio:APL:2009,Navio.NJP2010}
and also in biomedical
applications~\cite{2018_Chemistry_Materials_polyN_FeN_theory}.
However, the fundamental understanding of mononitride FeN compound
is still not well understood and variances can be seen between
theory and experiments in terms of the lattice parameter (LP),
structure and the magnetic ground state.

Considering recent theoretical works
~\cite{2018_Chemistry_Materials_polyN_FeN_theory,2019_JAC_FeN_1st_pricipal,2017_JPCS_FeN_CoN_DFT,2015_RSC_ADV_mag_order,2014_JAC_3d_mononitride_1st_princ,2012_JAC_Soni,FeN:PRB:Houari}
and comparing with experimental works, various possibilities
emerge about the structure and the magnetic ground state of FeN
(i)~\gd-FeN with zinc-blende (ZB)-type structure
(LP$\sim$4.3\,{\AA}) and a nonmagnetic (NM) ground state
(ii)~\gt-FeN with rock-salt (RS)-type structure
(LP$\sim$4.5\,{\AA}) and a ferromagnetic (FM) or antiferromagnetic
(AFM) ground state (iii) NiAs-type structure with a FM ground
state (iv) CsCl-type structure with NM ground state (iv)
wurtzite-type structure with a NM ground state (v) MnP-type
structure at very high pressure. Among these, FeN in CsCl,
wurtzite and MnP structures has not yet been synthesized
experimentally, NiAs-type structure was experientially evidenced
recently under
HPHT~\cite{2018_inorgchem_FeN2,2018_Nat_Com_poly_FeNx,2017_Iorg_Chem_FeN_FeN2,2018_JAC_FeN_megabar,2017_Angew_Chem_NiAs_FeN}.
Under ambient temperature and pressure conditions only the ZB and
RS-type FeN have been evidenced hitherto, however the existence of
RS-type FeN is still debated experimentally. Recent
first-principles calculations carried out by Li et al.
~\cite{2019_JAC_FeN_1st_pricipal} predict that the total energy
for ZB-FeN (non spin polarized) at -16.893\,eV is only slightly
lower than for RS-FeN (spin polarized) at -16.892\,eV. This energy
is significantly higher for all other FeN phases mentioned above.
Therefore, at T = 0\,K and P = 0\,GPa, the preferred phase is
ZB-type FeN followed by RS-type FeN.


From a review of recent experimental works probing the magnetic
ground state of FeN, it can be seen that irrespective of its LP
(4.3 or 4.5\,{\AA}), FeN remains non-magnetic down to 2\,K. An
exception to this, are studies peformed by Usui and
Yamada~\cite{2012_HI_FeN_PLD,2013_HI_FeN_PLD} and earlier works of
Nakagawa~\cite{Nakagawa.HypInt.1991}, Suzuki~\cite{Suzuki.JAC1993}
and Hinomura~\cite{1997_PhysicaB_Hinomura} et al. In all these
works a magnetic phase was found at low temperature but along with
an oxide phase. Usui and
Yamada~\cite{2012_HI_FeN_PLD,2013_HI_FeN_PLD} deposited FeN films
using PLD and by varying the nitrogen partial pressure they
synthesized ZB-type \gd-FeN and RS-type \gt-FeN. In these works
low temperature (down to 5\,K) \iFe~M\"{o}ssbauer spectroscopy
measurements were carried out and it was concluded that \gd-FeN
remain NM and \gt-FeN become AFM at low temperature. The N\'{e}el
temperature (\Tn) estimated for \gt-FeN films was 220\,K. In
addition to the AFM component with a hyperfine field of about
30\,T an additional component with hyperfine field of 49\,T was
also observed. Generally, such component stems from an oxide
phase~\cite{Nakagawa.HypInt.1991,1997_PhysicaB_Hinomura}
therefore, the purity of these samples was questionable.

Jouanny et al.~\cite{Jouanny2010TSF} prepared FeN thin films
having LP $\sim$4.5\,{\AA} using sputtering at 373\,K and from
their zero-field-cooled and field-cooled (1\,T) magnetization
measurements they could not find the presence of any magnetic
order. J. Pak et al.~\cite{JPak_2010} studied the growth behavior
of epitaxial FeN thin films deposited using N plasma assisted MBE
and from their \textit{in-situ} reflection high energy electron
diffraction (RHEED) measurements they found that the LP of
ultrathin FeN films was about 4.52\,{\AA}. Up to about 53
monolayers (ML), the LP of FeN films remained at this value but
for a 140 ML thick ($\sim$60\,nm) FeN film, the LP reduced to
4.32\,{\AA}. This reduction in LP was explained in terms of an
intrinsic in-plane tensile stress that generates during the growth
of the film. They also did magnetization and susceptibility
measurements on the 60\,nm thick sample down to 5\,K and could not
find the presence of any magnetic order. Gupta et
al.~\cite{MG:JAC:2011} prepared FeN films with LP$\sim$4.5\,{\AA}
using sputtering at 300\,K and did \iFe~M\"{o}ssbauer spectroscopy
measurements down to 5\,K and could not find any magnetic ordering
even after applying a magnetic field of 5\,T. It may be noted that
the LP of FeN films seems to vary with the growth temperature
(\Ts), when \Ts $<$ 400\,K, LP $\sim$ 4.5\,{\AA} and between
400-500\,K its was LP$\sim$4.3\,K.

Above experimental results indicate that ultrathin FeN films may
behave differently than their thicker counterparts. When the
growth of FeN films starts their LP is large and an intrinsic
in-plane tensile stress forces it to reduce. In such a situation
it can be assumed that initial FeN layer grow with a RS-type motif
which changes to ZB-type when film thickness increases beyond a
critical value. This critical thickness can be changed by varying
the deposition conditions. However, to prove this point, low
temperature magnetic measurements needs to be done on ultrathin
films, preferably during \textit{in-situ} growth to avoid surface
contamination. Hitherto, low temperature magnetization
measurements were performed only on thick FeN films ($>$ 50\,nm).

Magnetization and structural measurements at low temperatures on
ultrathin films combined with \textit{in-situ} thin film growth is
a challenging task. In the present work we achieved this by
preparing \textit{in-situ} systems capable of doing x-ray
absorption near edge spectroscopy (XANES) measurements at N
$K$-edge and nuclear resonant scattering (NRS) measurements. It is
known that N $K$-edge XANES can provide reliable information about
the coordination between metal and N
atoms~\cite{1997_SSR_JG_Chen_Review} and NRS is a technique which
is analogues to M\"{o}ssbauer spectroscopy thereby providing the
information about the local magnetic
structure~\cite{1996_HI_Ruffer_NRS}. In both cases FeN films were
grown using a similar magnetron sputtering source ($\phi$1\,inch)
and under similar deposition conditions. From N $K$-edge XANES
measurements, we observed a structural transition taking place
between 5-10\,nm and low temperature (100\,K) NRS measurements
also exhibited the presence of a magnetic component at similar
thicknesses. These first ever \textit{in-situ} N $K$-edge and NRS
measurements clearly reveal that ultrathin FeN films grow in the
RS-type structure and are magnetic at 100\,K. When the thickness
of FeN film increases beyond few nm, the structure changes to
ZB-type FeN which has a non-magnetic ground state.

\section{Experimental}
\label{exp}

\begin{figure*}
\includegraphics[width=0.5\textwidth]{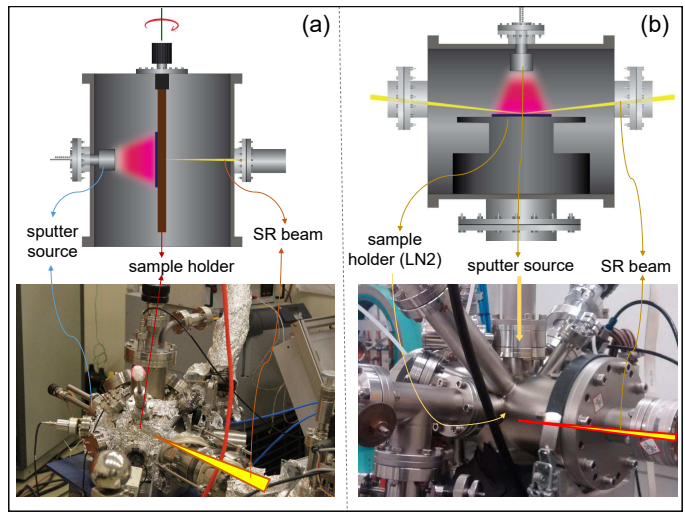}
\caption{A schematic diagram and a photograph of the in-situ
system prepared at the soft x-ray absorption spectroscopy beamline
BL01 beamline at Indus 2 (a) and at the nuclear resonance
scattering beamline P01 at Petra III (b). In both cases a
$\phi$1\,inch magnetron source was used.}
\label{fig:1}       
\end{figure*}
In order to study the early stages of film growth, a magnetron
sputtering system was mounted separately at the soft x-ray
absorption spectroscopy (SXAS) beamline BL01~\cite{XAS_beamline}
of Indus 2 synchrotron radiation (SR) source at RRCAT, Indore,
India and at the NRS beamline P01~\cite{p01c} of Petra III SR
radiation source at DESY, Hamburg, Germany. In both cases a
$\phi$1\,inch magnetron source (AJA Int. Inc.) with Fe or
\iFe~target was used and keeping experimental conditions similar,
FeN films of thickness ranging from sub nm to few tens of nm were
deposited. Nitrogen alone was used to sputter the iron target. By
using nitrogen alone as the sputtering medium (instead of a
mixture of Ar+N$_2$), single phase FeN films can be synthesized in
reactive dc magnetron
sputtering(dcMS)~\cite{Nakagawa.HypInt.1991,MG:JAC:2011,2015:PRB:MG:FeN,2019_JMMM_Fe4N_by_FeN}.
After growing a film of a particular thickness, XANES or NRS
measurements were performed. Prior to \textit{in-situ}
experiments, few test samples were deposited to determine the
deposition rates (using x-ray reflectivity), the crystalline
structure using x-ray diffraction (XRD, Bruker D8 Advance using
Cu-$K_{\alpha}$ x-rays) and the magnetic structure using room
temperature conversion electron M\"{o}ssbauer spectroscopy (CEMS).
The details of the experimental stations exclusively prepared in
this work are described below.

\subsection{\textit{In-situ} system at SXAS beamline}
\label{exp:sxas}

In the ultra-high vacuum (UHV) experimental station of SXAS
beamline, a magnetron sputtering source was installed. This source
was mounted as shown in the schematic diagram in fig.~\ref{fig:1}
(a). A photograph of the experimental chamber with the sputter
source is also shown in fig.~\ref{fig:1} (a). Using this source
FeN films were deposited on a Si substrate facing the source and
then the substrate was rotated along its axis so as to allow the
SR x-rays to fall directly on to the deposited sample. A mass flow
controller was also mounted to control the gas flow precisely.
Using N$_2$ (purity 99.9995\%) alone as the sputtering medium, an
iron target (purity 99.995\%) - $\phi$1\,inch and thickness
0.5\,mm was sputtered. The base pressure in the vacuum chamber was
about 2$\times$10$^{-8}$\,mbar and with N$_2$ gas flowing at about
20\,sccm, the pressure during deposition was about
5$\times$10$^{-3}$\,mbar.

The sputter source was equipped with a shutter so that the actual
film deposition can be started after stabilizing the source
conditions. Prior to in-situ experiments, Fe target was
pre-sputtered to remove surface contaminations. The sputtering
power used was about 15\,W (450\,V dc) and the target - substrate
distance was about 5\,cm. All samples were deposited at ambient
temperature without any intentional heating on a clean Si
substrate. After each deposition the chamber was evacuated back to
UHV conditions and XANES measurements at the N $K$-edges were
carried out in the total electron yield mode. Typically, a N
$K$-edge scan takes about a couple of minutes and a number of
scans were averaged to improve the data reliability and
statistics. The energy resolution at the N $K$-edge is better than
0.2\,eV. The effectiveness of such \textit{in-situ} growth using
SXAS has been amply demonstrated during the growth of
Cr~\cite{2010_JAC_Cr_XAS_insitu},
TiO$_2$~\cite{2014_BJN_TiO2_insitu_XAS},
TiN~\cite{2017_AIP_TiN_RG} and very recently for Fe films on
MgO~\cite{2019_JPCM_Fe_insitu}.

\subsection{\textit{In-situ} system at NRS beamline}
\label{exp:nrs}

Similarly for \textit{in-situ} NRS measurements also a
$\phi$1\,inch MS source was installed in a vacuum chamber as shown
in the schematic diagram and photograph in fig.\ref{fig:1} (b).
Here instead of natural Fe, a $^{57}$Fe target with about 95\%
abundance and 99.95\% purity was used. The thickness of this
target was 0.4\,mm. The sample holder was either held at ambient
temperature (300\,K) or cooled with liquid nitrogen (LN$_2$)
flowing continuously in the copper reservoir holder. The
temperature achieved at the substrate surface was about 100\,K.
The base pressure achieved in the chamber was about
3$\times$10$^{-8}$\,mbar (after a bake out). All other deposition
conditions were kept similar as described in
section~\ref{exp:sxas}. Samples were deposited on a sapphire
substrate.

Here also after depositing a film, the chamber was evacuated again
to UHV conditions and then NRS and x-ray (electronic) reflectivity
(XRR) was measured. Since both NRS and XRR measurements can be
done, almost simultaneously (one after the other within few
minutes), the information about the magnetic structure and sample
thickness, density, roughness etc. can be obtained in a unique
way. The SR beam was tuned to $^{57}$Fe nuclear resonance energy
of 14.4\,keV. Utilizing the pulse structure of SR beam, the prompt
scattering taking place within few ns was used to record the XRR
pattern, in the $\theta$-2$\theta$ mode and NRS was recorded using
a multi channel analyzer between 5 to 150\,ns in the forward
scattering geometry. The angle of incidence during NRS
measurements was just above the critical angle. The details of the
beamline can be in found in \cite{p01c,wille2010NFS} and about the
NRS of SR in
\cite{1996_HI_Ruffer_NRS,2013_Kai_NRS,1996_HI_Smirnov_NRS_SR,2008_CRP_NRS_Ruffer,Ralf_NRS}.

\section{Results and Discussion}
\label{rnd}

\subsection{Structural and magnetic characterization of FeN thin film}
\label{rnd:char}

\begin{figure} \vspace{-9mm} \center
\includegraphics [width=0.45\textwidth] {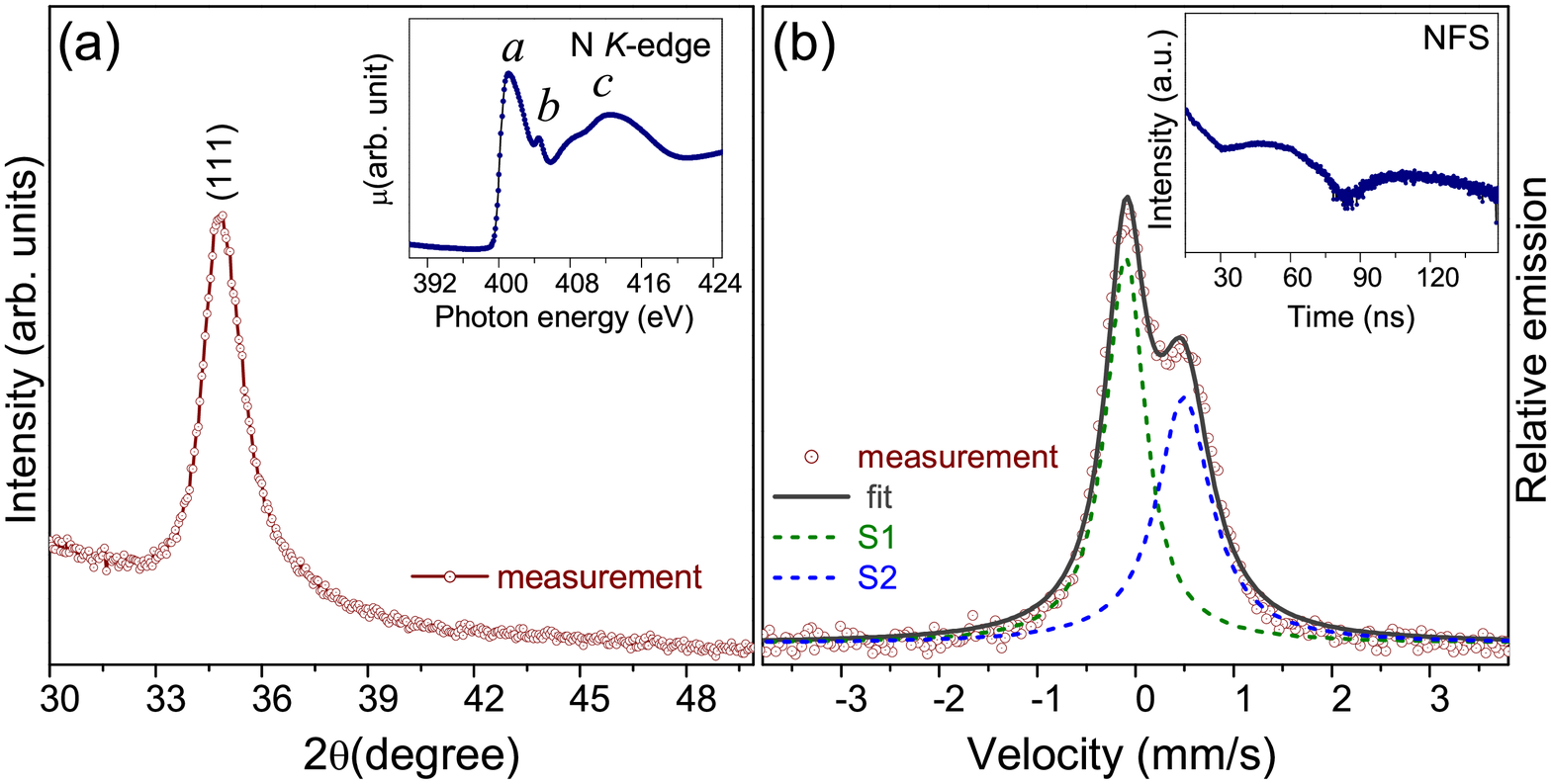} \vspace{-18mm}
\caption{X-ray diffraction pattern of a 25 and 50\,nm FeN film
grown at 300\,K (a) and the inset shows the N K-edge XANES pattern
of 50\,nm sample. Conversion electron M\"{o}ssbauer spectrum of a
50\,nm $^{57}$FeN thin film (b) and the NRS pattern of this sample
is shown in the inset of (b).}
\label{fig:xrd_xas}       
\end{figure}

Prior to in-situ experiments, preliminary characterization of a
50\,nm thick FeN samples was carried out to understand its
structural and magnetic properties. Shown in fig.\ref{fig:xrd_xas}
(a) is the XRD pattern of this sample exhibiting a broad peak
around 2$\theta$ = 34.9$^{\circ}$ with a grain size of about
12.5($\pm$0.5)\,nm. This peak can be assigned to FeN(111) and the
LP comes out to be 4.45\,{\AA}. Generally, the LP of FeN films has
been found anywhere between 4.3 to 4.55\,{\AA} and when FeN films
are deposited at low \Ts~($<$ 400\,K), LP$\sim$
4.5\,{\AA}~\cite{Jouanny2010TSF,MG:JAC:2011,Naito:FeN:14} and at
high \Ts, LP reduces to about
4.3\,{\AA}~\cite{JPak_2010,Naito:FeN:14}. Therefore, from our XRD
measurements, we can confirm the formation of mononitride FeN
phase.

The N $K$-edge XANES pattern of this test sample is shown in the
inset fig.\ref{fig:xrd_xas} (a). Here, prominent features are
labelled as $a$, $b$ and $c$. Here $a$ and $b$ correspond to
transition from N 1$s$ to hybridized states of Fe 3$d$ and N 2$p$.
Features $a$ and $b$ can be assigned as t$_{2g}$ and e$_g$ in the
octahedral and e and t$_2$ in the tetrahedral symmetry as
explained later in section~\ref{insitu:sxas}. Higher energy
feature $c$ is due to electronic transition from N1$s$ to a
hybridized N2$p$ and Fe4$sp$
states~\cite{1997_SSR_JG_Chen_Review}. Although N $K$-edge XANES
has been used as a finger print to identify the electronic
structure of early transition metal nitrides, such measurements in
FeN system are almost nonexistent. In our recent work, we found
that the intensity of feature $b$ becomes almost negligible
(compared to feature $a$)~\cite{2015:PRB:MG:FeN} in 150-200\,nm
thick FeN films deposited at 300\,K. Generally, in octahedral
symmetry both $a$ and $b$ features are clearly resolved e.g. in
case of TiN~\cite{1993_JESRP_SORIANO_TiN_XAS} and
CrN~\cite{2004_SSC_N_K_edge_CrN_C2N,2019_TSF_CrN_MG}. In fact N
$K$-edge features have been compared in fcc CrN and hcp Cr$_2$N
and it was anticipated that due to octahedral crystal field, N
$K$-edge features show splitting in CrN but in Cr$_2$N, no such
splinting can be seen in absence of octahedral crystal
field~~\cite{2004_SSC_N_K_edge_CrN_C2N}. Therefore, N $K$-edge
XANES can be used to distinguish between ZB and RS-type FeN, as
demonstrated from our \textit{in-situ} measurements presented in
section~\ref{insitu:sxas}.

The magnetic state of another test sample (50\,nm \iFe{N} film)
was measured using CEMS as shown in fig.~\ref{fig:xrd_xas} (b). It
shows an asymmetric peak typically found in FeN films. This peak
was deconvoluted into two singlets - S1 and S2. We find that the
isomer shift of S1 is around 0, while that of S2 $\sim$ 0.5\,mm/s.
Here, S1 can be assigned to Fe coordinated with four N atoms as
expected in ZB-type FeN; S2 can be understood to originate due to
various factors such as defects or vacancies in ZB-FeN in
agreement with previous
works~\cite{MG:JAC:2011,Jouanny2010TSF,2006_NIMB_stoichiometic_FeN,2003_HI_Borsa}.
The NFS pattern of this sample is shown in the inset of
fig.~\ref{fig:xrd_xas} (b). The featureless pattern confirms that
this sample is non-magnetic. Details of NFS results are discussed
in section~\ref{rnd:nfs100K}. Therefore, our 50\,nm sample can be
assigned as non-magnetic FeN (at 300\,K) having a ZB-type
structure with a LP of 4.45\,{\AA}.

\subsection{\textit{In-situ} NRS measurements at 300\,K}
\label{rnd:nfs300K}

\begin{figure} \center
\includegraphics [width=0.45\textwidth] {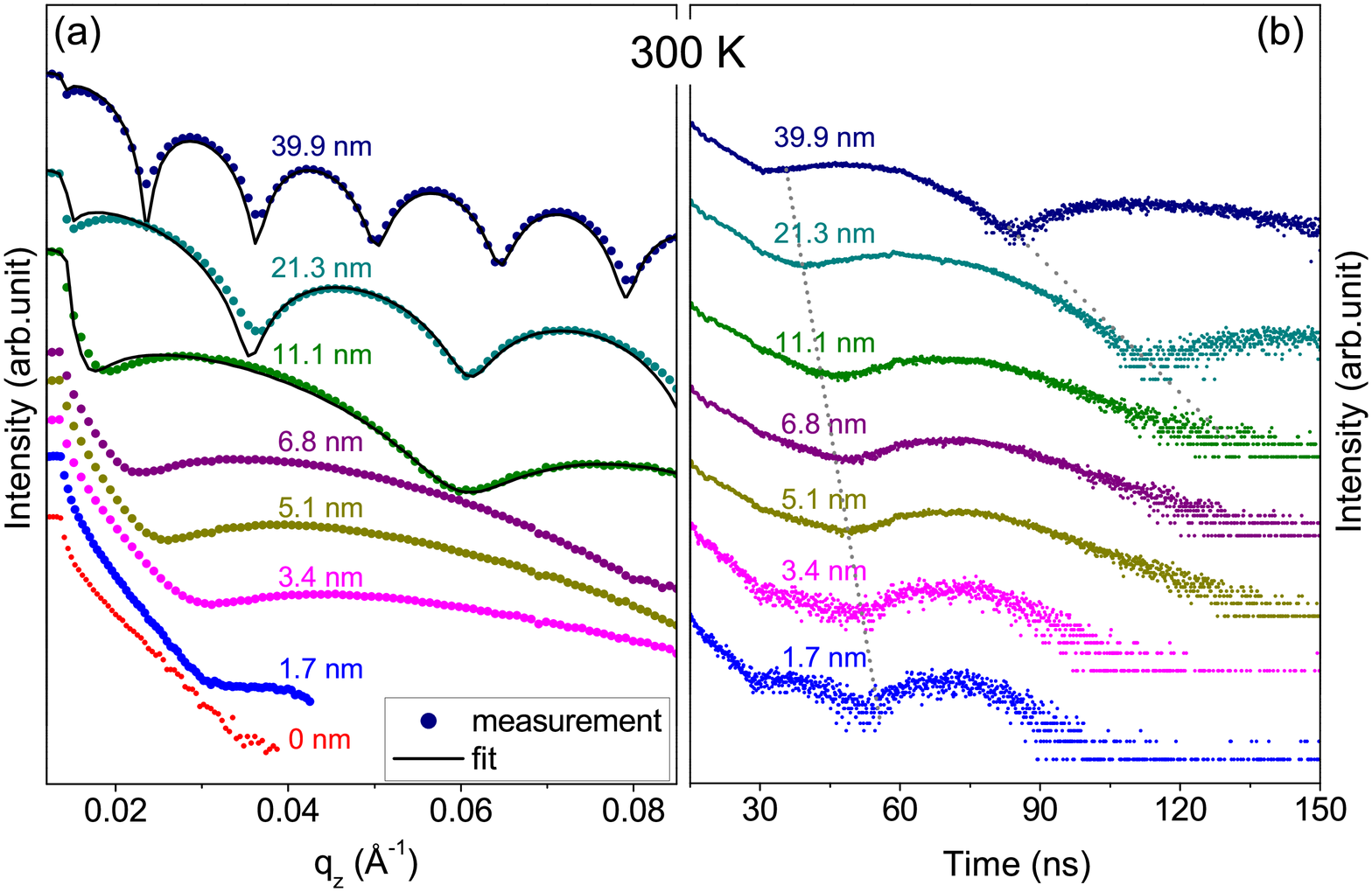}
\vspace{-1mm} \caption{X-ray reflectivity (a) and NRS pattern (b)
of \iFe{N} thin film grown \textit{in-situ} on a sapphire
substrate at room temperature (300\,K). Measurements have been
carried out using 14.4\,keV x-rays.}
\label{insitu:nfs:300K}       
\end{figure}

Figure~\ref{insitu:nfs:300K}(a,b) shows the XRR and NRS patterns
taken during \textit{in-situ} deposition of \iFe{N} thin film for
various film thicknesses. As films become thick, the frequency of
oscillation increases in the XRR pattern signifying that the
thickness of films is increasing. The fitting of XRR pattern
reveal that film roughness is within 1-2\,nm and estimated film
density matches well with FeN films. The NRS pattern of all FeN
films show broad oscillations - marked by a dashed line as shown
in fig.~\ref{insitu:nfs:300K}(b). These broad oscillations arise
due to thickness of the film and are known as dynamical beats,
explained in section~\ref{rnd:nfs100K}. The absence of beating
patterns arising due to hyperfine interactions (electric
quadrupole or magnetic dipole) signifies that all FeN films are
non-magnetic, irrespective of film thickness. Only the thinnest
sample (1.7\,nm) shows an additional beat which could have
magnetic origin but due to low thickness and poor statistics, it
is difficult to assign any magnetic component. FeN films at 300\,K
are expected to be non-magnetic. They might show a magnetic
signature when cooled below 220\,K~\cite{2013_HI_FeN_PLD}.
Therefore, our 300\,K \textit{in-situ} measurements demonstrate
the capability of the technique to probe ultrathin films. In order
to probe the magnetic ground state we did \textit{in-situ} XRR and
NRS measurements at 100\,K which are presented in
section~\ref{rnd:nfs100K}.

\subsection{\textit{In-situ} N $K$-edge XANES measurements}
\label{insitu:sxas}

\begin{figure*} \center
\includegraphics [width=0.6\textwidth] {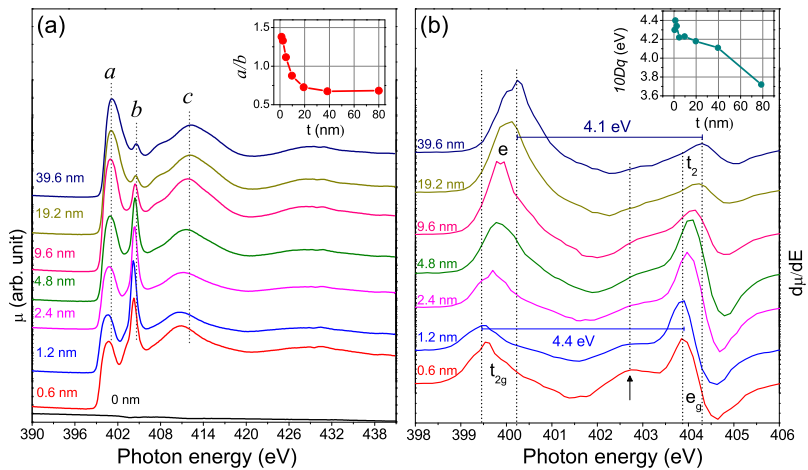}
\vspace{-1mm} \caption{N $K$-edge XANES measurements performed
\textit{in-situ} during the growth of FeN film at 300\,K using
reactive magnetron sputtering. Prominent N $K$-edge features can
also be seen in sub nm FeN film.}
\label{fig:insitu_xas}       
\end{figure*}

Figure~\ref{fig:insitu_xas} shows the N $K$-edge XANES patterns
taken during the \textit{in-situ} growth of FeN films at 300\,K.
Before starting the deposition, the chamber was flushed with N$_2$
gas 2-3 times and after evacuating the chamber to UHV conditions,
N $K$-edge measurements were performed on the bare substrate. As
expected no signatures of N $K$-edge features can be seen on the
bare substrate. Subsequently FeN film was deposited for a short
time and it was astonishing to see that prominent N $K$-edge
features can be seen even when film thickness was less than a
monolayer (not shown) and for a 0.6\,nm film, well-resolved N
$K$-edge features can be seen in fig.~\ref{fig:insitu_xas}(a). The
assignment of these features as $a$, $b$ and $c$ was already
explained in sec.~\ref{rnd:char}. Shown in
fig.~\ref{fig:insitu_xas}(b) is a magnified view of the derivative
of absorption coefficient (d$\mu$/dE) depicting the precise
positions of features $a$ and $b$. For ultrathin films ($<$
5\,nm), we can see that the feature $b$ is the strongest, also it
shows a shoulder on the lower energy side (marked by an arrow in
fig.~\ref{fig:insitu_xas}(b)), which could be due to some
splitting in the e$g$ orbitals.

When the film thickness increases to 9.6\,nm, this shoulder
disappears and the feature $b$ suddenly becomes the weakest. For
films of even higher thickness the feature $b$ becomes even
smaller (than both $a$ and $c$). The ratio of features $a$ and $b$
is plotted in the inset of fig.~\ref{fig:insitu_xas}(a). Another
prominent change that can be clearly seen is the shift in the
position of features $a$ and $b$. We can see that both $a$ and $b$
shift to higher energy as thickness increases and the energy
difference of these two features ($10Dq$) reduces from 4.4\,eV to
4.1\,eV when the thickness increases from 1.2 to 39.6\,nm. This
difference becomes even smaller ($\sim$3.7\,eV) when the film
thickness become even larger at about 80\,nm (not shown). A
variation in the energy separation $10Dq$ as a function of film
thickness is shown in the inset of ~\ref{fig:insitu_xas}(b).

The intensity ratio and the energy separation of the first two
features of ligand $K$-edge (e.g. O or N) in a transition metal
oxide or nitride can be used to probe the hybridization between
metal 3$d$ and N or O 2$p$
orbitals~\cite{1997_SSR_JG_Chen_Review}. The energy separation
$10Dq$ is a direct measure of ligand-field splitting and as atomic
number increases $d$-orbitals become more extended resulting in
larger overlap with 2$p$ orbitals and therefore $10Dq$ increases.
It may be noted that in N $K$-edge XANES of TiN is well-studied
and typically, $10Dq$ is about
2.5\,eV~\cite{2017_AIP_TiN_RG,1993_JESRP_SORIANO_TiN_XAS} and in
thick FeN films it comes around 3.7\,eV, this larger value of
$10Dq$ in FeN (as compared to TiN) can be understood in terms
extended $d$ orbitals in FeN as compared to TiN. Even larger value
of $10Dq$ $\sim$ 4.4\,eV was observed in ultrathin FeN films. It
can be understood if we assume that the symmetry of N atoms
changes from tetrahedral (ZB-type) to octahedral (RS-type). Due to
crystal field splitting five degenerate 3d orbitals split into
t$_{2g}$ and e$_g$ orbitals. In an octahedral symmetry, the three
t$_{2g}$ orbitals points towards the corner and the two e$_g$
orbitals towards the center of the cube. Therefore, the
interaction of e$_g$ orbitals with a ligand (N) is stronger than
those of t$_{2g}$ orbitals. On the other hand, in a tetrahedral
symmetry, the crystal field splitting is reversed, and $e$
orbitals are lower in energy than those of t$_2$ orbitals. Since
in case of tetrahedral coordination the symmetry is
non-centrosymmetric, the gerade ($g$) label is removed. Therefore,
the reduction in $10Dq$ with an increase in film thickness can be
understood if the structure changes from RS-type FeN in ultrathin
films to than in ZB-type FeN in thicker films.

On the other hand, variations in the intensity ratios of first two
features $a$ and $b$ are even more prominent across the films
thickness, in particularly between film thickness of 1 to 10\,nm.
The structural transition from RS-type to ZB-type could be one
factor that influences the relative intensity of features $a$ and
$b$, but the sharpness of the feature $b$ (FWHM $\sim$ 0.8\,eV)
and $b$ $>>$ $a$, is somewhat puzzling. Such an intense feature in
N $K$-edge has been observed during the adsorption of nitrogen on
Mo(110) surface and has been explained in terms of N$-$N single
and N$\equiv$N triple bonds that are formed on the surface of
Mo~\cite{1997_SSR_JG_Chen_Review}. Polarization-dependent
measurements of N $K$-edge at the normal and grazing incidence by
Chen et al.~\cite{1997_SSR_JG_Chen_Review} evidenced a similar
sharp feature resulting due to formation of N$-$N single bond
perpendicular to the surface. The presence of such sharp N
$K$-edge feature can also be seen during oxidation of
TiN~\cite{1993_JESRP_SORIANO_TiN_XAS} and also after the
implantation of N$^+$ ions on Zr and
Hf~\cite{1993_JESRP_SORIANO_TiN_XAS,1997_SSR_JG_Chen_Review}. In
fact such N$-$N or N$\equiv$N bonds are also expected to be
present as polymeric N chains in N rich \FeNx(x $>$ 1), evidenced
recently under HPHT by Bykov et al.~\cite{2018_Nat_Com_poly_FeNx}.
Recently Wu et
al.~\cite{2018_Chemistry_Materials_polyN_FeN_theory} predicted the
existence of FeN$_4$ phase even under ambient pressure and
temperature. The occurrence of such a sharp N $K$-edge feature may
also stem from the presence of polymeric N chains in ultrathin FeN
films but this needs to be further confirmed. The observed changes
in the N $K$-edge XANES clearly demonstrate that ultrathin FeN
films ($ < $ 5\,nm) show a completely different behavior than
rather thick FeN films and indicate that a structural transition
from RS-type FeN to ZB-type FeN is taking place in the thickness
range of 5 - 10\,nm.

\subsection{\textit{In-situ} NRS measurements at 100\,K}
\label{rnd:nfs100K}

From our room temperature (300\,K) \textit{in-situ} NRS, XRR and N
$K$-edge XANES measurements, the information obtained can be
summarized as (i) irrespective of film thickness, FeN films were
non-magnetic at 300\,K (ii) the N $K$-edge XANES pattern of
ultrathin films ($<$ 5\,nm) show a different characteristics
indicating that films undergo a structural transition from RS to
ZB-type as thickness increases beyond 5-10\,nm. As evidenced by
Yamada et al.~\cite{2013_HI_FeN_PLD}, RS-type FeN has a AFM ground
state with \Tn $\sim$ 220\,K, therefore by doing \textit{in-situ}
NRS measurements at low temperatures (below 220\,K) it should be
possible to verify if films become magnetic as expected for FeN in
the RS-type structure.

\begin{figure} \center
\includegraphics [width=0.45\textwidth] {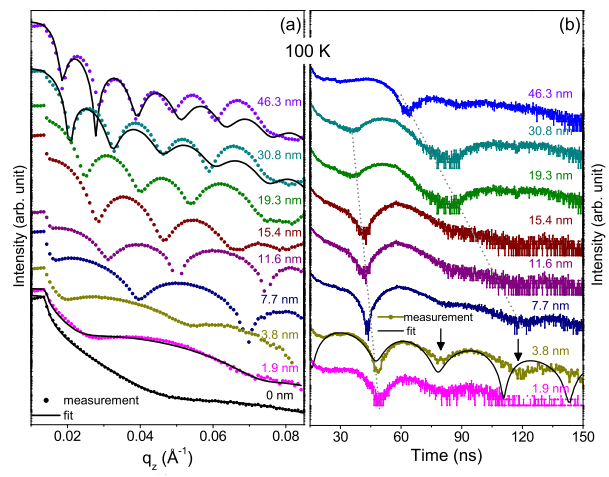}
\vspace{-1mm} \caption{X-ray reflectivity (a) and NRS pattern (b)
of \iFe{N} thin film grown \textit{in-situ} on a sapphire
substrate at 100\,K.}
\label{fig:nfs100K}       
\end{figure}

Similar to the procedures followed for \textit{in-situ} NRS
measurements at 300\,K, the growth of FeN films was also studied
by cooling the substrate holder by a continuous flow of LN$_2$.
Prior to cooling the sample holder, the whole system was baked to
achieve UHV conditions which is an essential requirement to avoid
condensation of moisture on the substrate surface at LN$_2$
temperatures. The estimated temperature at the sample surface was
about 100\,K and films of different thicknesses were deposited at
this temperature. Figure~\ref{fig:nfs100K}(a) and (b) shows XRR
and NRS patterns for FeN films of different thicknesses. Here
again total thickness oscillations can be seen in the XRR pattern
(fig.~\ref{fig:nfs100K}(a)) yielding the thickness of films. The
deposition rate estimated from the known deposition time was about
7.7\,nm/min which was more than twice than that of samples
deposited at 300\,K (3.4\,nm/min). Such enhancement in deposition
rates is not unexpected at LN$_2$ temperatures as evidenced
recently during the growth of nickel oxide thin
films~\cite{2018_JVST_NiO_LT}. From the fitting of XRR patterns,
we found that the density of films deposited at 100\,K was similar
to those at 300\,K. More interesting behavior can be seen in the
NRS patterns shown in fig.~\ref{fig:nfs100K}(b). Here broad
oscillation correspond to dynamical
beats(DBs)~\cite{PRB_DB_1,PRB_DB_2} originating due to the total
thickness of sample (marked by dashed lines) can be seen. With an
increase in thickness DBs shift to lower time in agreement with
the behavior that can also be seen in samples deposited at 300\,K
(fig.~\ref{insitu:nfs:300K}(b)).

On the other hand, an additional beating pattern can be seen in
ultrathin films ($\leq$ 7.7\,nm) and more prominently for the
3.8\,nm film. It is known that in the presence of hyperfine
interactions due to electric quadrupole or magnetic dipole,
nuclear levels show quadrupole or Zeeman splitting. During the
nuclei de-excitation process interference amplitudes give rise to
oscillations in the scattered intensity with time. These
oscillations are known as quantum beats (QBs)~\cite{PRL_QB_1}. The
periods and amplitudes of QBs reveal the information about the
magnetic structure present in the sample. The absence of QBs in
our NRS pattern is a characteristic of a non-magnetic sample.
These QBs are quite clearly visible in the 3.8\,nm sample, and
their presence, though weak, can also be seen in the 1.9 and
7.7\,nm films. The positions of QBs is marked by an arrow in
fig~\ref{fig:nfs100K}(b). The presence of QBs indicates that the
sample possess a magnetic structure. In order to find the nature
of magnetism, the NRS spectra was fitted using REFTIM
software~\cite{REFTIM2008}. Assuming a hyperfine field of about
15\,T, the the period of QBs matches well with our experimental
data. This value is smaller than the hyperfine filed of 30\,T
expected for the AFM RS-type
FeN~\cite{2013_HI_FeN_PLD,2015_RSC_ADV_mag_order}. Nevertheless,
the presence of magnetic state in our ultrathin FeN films confirm
that RS-type FeN indeed possess a magnetic ground state. Since
samples in our study have been deposited at about 100\,K, it can
be anticipated that crystalline ordering may not get established
due to lack of adatom mobility and therefore in this situation
structural disorders may lead to a smaller hyperfine filed
evidenced in the 3.8\,nm film. Ideally, by depositing the film at
a higher substrate temperature around 450\,K and then cooling it
down to low temperature, precise determination of the magnetic
structure can be made.

A close correlation can be drawn between the N $K$-edge XANES and
the 100\,K NRS pattern of ultrathin ($<$ 5\,nm) FeN thin films. At
such a low thickness their $10Dq$ value is large (4.4\,eV), the
ratio of features $b/a$ is also large and they are
antiferromagnetic. The octahedral coordination between Fe and N
atoms confirm the RS-type structure and the presence of AFM ground
state further confirms it. From the observed growth behavior, it
appears that initially FeN films grow in the RS-type structure,
subsequently due to an in plane tensile stress, films are forced
to adopt the ZB-type structure. Such transformation in the
structure was also evidenced by J. Pak et al.~\cite{JPak_2010}
during the \textit{in-situ} growth and study of epitaxial FeN
films. It was found that the LP was 4.52\,{\AA} in the beginning
and when the thickness increases to about 60\,nm it reduces to
4.32\,{\AA}. It seems that the critical thickness at which such
transition takes place will depend upon the growth temperature, in
our case films are grown at 300\,K and the RS to ZB-type
structural transition is taking place around 5\,nm but when films
were grown at high temperature (e.g. 483\,K by J. Pak. et al.),
such transition was taking place at a much larger thickness. It is
anticipated that by growing FeN films at high substrate
temperature and then by studying the magnetic properties of
ultrathin films at low temperature more precise estimation of
magnetic ground state of FeN can be made.

\section{Conclusions}

In this work structure and magnetic properties of ultrathin FeN
films were studied during \textit{in-situ} growth using N $K$-edge
XANES at 300\,K and NRS measurements at 300 and 100\,K. N $K$-edge
XANES measurements clearly reveal that coordination between Fe and
N atoms changes from octahedral (RS-type FeN) to tetrahedral
(ZB-type FeN) when thickness increases beyond 5\,nm. Low
temperature NRS measurements also depict such behavior as RS-type
FeN films show an antiferromagnetic character. When the thickness
of films increases beyond 5\,nm films become non-magnetic as
expected in for ZB-type FeN. The remarkable sensitivity of both
XANES and NRS for ultrathin layers can further be utilized by
growing FeN films at high temperatures and then studying the
structural and magnetic properties at low temperatures.

\begin{acknowledgements}
We would like to thank to the Department of Science and
Technology, India for the financial support and Jawaharlal Nehru
Centre for Advanced Scientific Research (JNCASR) for managing the
project. We acknowledge technical help received from L. Behera, A.
Gome and R. Sah. We are thankful to V. Ganesan and A. K. Sinha for
support and encouragements.

\end{acknowledgements}

\bibliographystyle{spphys}       


\begin{thebibliography}{10}
\providecommand{\url}[1]{{#1}} \providecommand{\urlprefix}{URL }
\expandafter\ifx\csname urlstyle\endcsname\relax
  \providecommand{\doi}[1]{DOI \discretionary{}{}{}#1}\else
  \providecommand{\doi}{DOI \discretionary{}{}{}\begingroup
  \urlstyle{rm}\Url}\fi

\bibitem{Bhattacharyya_FeN_Review}
S.~Bhattacharyya, The Journal of Physical Chemistry C
\textbf{119}, 1601
  (2015).


\bibitem{2000_JMMM_Fe16N2_gmm_or_not}
M.~Takahashi, H.~Shoji, J. Magnetism and Magnetic Materials
\textbf{208},
  145  (2000).


\bibitem{2016_Sci_Rep_Fe16N2}
Y.~Jiang, M.A. Mehedi, E.~Fu, Y.~Wang, L.F. Allard, J.P. Wang,
Scientific
  Reports \textbf{6}, 25436 (2016).


\bibitem{2006_PRB_Fe4N_SPR}
S.~Kokado, N.~Fujima, K.~Harigaya, H.~Shimizu, A.~Sakuma, Phys.
Rev. B
  \textbf{73}, 172410 (2006).


\bibitem{JMMM10_Matar}
A.~Houari, S.F. Matar, M.A. Belkhir, Journal of Magnetism and
Magnetic
  Materials \textbf{322}, 658  (2010).


\bibitem{APL11_Ito_Fe4N}
K.~Ito, G.H. Lee, K.~Harada, M.~Suzuno, T.~Suemasu, Y.~Takeda,
Y.~Saitoh,
  M.~Ye, A.~Kimura, H.~Akinaga, Applied Physics Letters \textbf{98}, 102507
  (2011).

\bibitem{2009_PSS_B_Fe4N_th_exp_Blanca}
E.L.P.y. Blanc\'{a}, J.~Desimoni, N.E. Christensen, H.~Emmerich,
S.~Cottenier,
  physica status solidi (b) \textbf{246}, 909 (2009).


\bibitem{2018_APL_Fe4N_underlayer}
H.~Li, X.~Li, D.~Kim, G.~Zhao, D.~Zhang, Z.~Diao, T.~Chen, J.P.
Wang, Applied
  Physics Letters \textbf{112}, 162407 (2018).


\bibitem{2016_JAC_Fe3N_soft_mag}
W.~Yin, D.~Zhang, P.~Zhang, X.~Wang, W.~Wang, X.~Lei, Z.~Shi,
H.~Yang, Journal
  of Alloys and Compounds \textbf{688}, 828  (2016).


\bibitem{2017_nano_energy_Fe3N_li-on_battery}
H.~Huang, S.~Gao, A.M. Wu, K.~Cheng, X.N. Li, X.X. Gao, J.J. Zhao,
X.L. Dong,
  G.Z. Cao, Nano Energy \textbf{31}, 74  (2017).

\bibitem{Cai_2000}
Nuclear Instruments and Methods in Physics Research Section B:
Beam
  Interactions with Materials and Atoms \textbf{168}, 422  (2000).


\bibitem{2018_inorgchem_FeN2}
D.~Laniel, A.~Dewaele, G.~Garbarino, Inorganic Chemistry
\textbf{57}, 6245
  (2018).


\bibitem{Fe3N4:JAP02}
Y.N. Xu, P.~Rulis, W.Y. Ching, J. Appl. Phys. \textbf{91}, 7352
(2002).


\bibitem{2002_APL_Fe3N4}
W.Y. Ching, Y.N. Xu, P.~Rulis, Applied Physics Letters
\textbf{80}, 2904
  (2002).


\bibitem{2011_Chemistry_FeN2_prediction}
M.~Wessel, R.~Dronskowski, Chemistry – A European Journal
\textbf{17}, 2598.


\bibitem{2018_Nat_Com_poly_FeNx}
M.~Bykov, E.~Bykova, G.~Aprilis, K.~Glazyrin, E.~Koemets,
I.~Chuvashova,
  I.~Kupenko, C.~McCammon, M.~Mezouar, V.~Prakapenka, H.P. Liermann,
  F.~Tasn\'{a}di, A.V. Ponomareva, I.A. Abrikosov, N.~Dubrovinskaia,
  L.~Dubrovinsky, Nature Communications \textbf{9}, 2756 (2018).


\bibitem{2018_Chemistry_Materials_polyN_FeN_theory}
L.~Wu, R.~Tian, B.~Wan, H.~Liu, N.~Gong, P.~Chen, T.~Shen, Y.~Yao,
H.~Gou,
  F.~Gao, Chemistry of Materials \textbf{30}, 8476 (2018).


\bibitem{Jack1951}
K.H. Jack, Proceedings of the Royal Society of London. Series A.
Mathematical
  and Physical Sciences \textbf{208}, 200 (1951).


\bibitem{Oueldennaoua.SM.1985}
A.~Oueldennaoua, E.~Bauer-Grosse, M.~Foos, C.~Frantz, Scripta
Metall.
  \textbf{19}, 1503 (1985).

\bibitem{Nakagawa.HypInt.1991}
H.~Nakagawa, S.~Nasu, H.~Fujii, M.~Takahashi, F.~Kanamaru,
Hyperfine Interact.
  \textbf{69}, 455 (1991).


\bibitem{Suzuki.JAC1993}
K.~Suzuki, H.~Morita, T.~Kaneko, H.~Yoshida, H.~Fujimori, J.
Alloys Compds.
  \textbf{201}, 11  (1993).


\bibitem{1997_PhysicaB_Hinomura}
T.~Hinomura, S.~Nasu, Physica B: Condensed Matter
\textbf{237-238}, 557
  (1997).


\bibitem{SchaafHypInt95}
P.~Schaaf, C.~Illgner, M.~Niederdrenk, K.P. Lieb, Hyp. Int.
\textbf{95}, 199
  (1995).

\bibitem{Niederdrenk_JAC96}
M.~Niederdrenk, P.~Schaaf, K.P. Lieb, O.~Schulte, J. Alloys and
Compd.
  \textbf{237}, 81 (1996).

\bibitem{Rissanen_JAC98}
L.~Rissanen, M.~Neubauer, K.P. Lieb, P.~Schaaf, J. Alloys and
Compds.
  \textbf{274}, 74 (1998).

\bibitem{Schaaf.PMS.2002}
P.~Schaaf, Prog. Mater. Sci. \textbf{47}, 1  (2002).


\bibitem{2001_Vacuum_MG_FeN_IBS}
M.~Gupta, A.~Gupta, S.~Chaudhari, D. M.~Phase, V.~Ganesan, M.R.
Rao, T.~Shripathi,
  B.~Dasannacharya, Vacuum \textbf{60}, 395  (2001).


\bibitem{2003_ASS_FeN_dcMS}
X.~Wang, W.~Zheng, H.~Tian, S.~Yu, W.~Xu, S.~Meng, X.~He, J.~Han,
C.~Sun,
  B.~Tay, Appl. Sur. Sci. \textbf{220}, 30  (2003).


\bibitem{2004_JAC_FeN_plasma_gas_cond}
D.~Peng, T.~Hihara, K.~Sumiyama, Journal of Alloys and Compounds
  \textbf{377}, 207  (2004).


\bibitem{Gupta:PRB05}
R.~Gupta, M.~Gupta, Phys. Rev. B \textbf{72}, 024202 (2005).


\bibitem{2006_TSF_FeN_TEM_Demange}
V.~Demange, T.H. Loi, P.~Weisbecker, E.~Bauer-Grosse, Thin Solid
Films
  \textbf{494}, 184  (2006).


\bibitem{Jouanny2010TSF}
I.~Jouanny, P.~Weisbecker, V.~Demange, M.~Grafout\'{e},
O.~Pe{\~{n}}a,
  E.~Bauer-Grosse, Thin Solid Films \textbf{518}(8), 1883  (2010).


\bibitem{gupta:JAP2011}
M.~Gupta, A.~Tayal, A.~Gupta, R.~Gupta, J.~Stahn, M.~Horisberger,
A.~Wildes, J.
  Appl. Phys. \textbf{110}, 123518 (2011).


\bibitem{MG:JAC:2011}
M.~Gupta, A.~Tayal, A.~Gupta, V.~Reddy, M.~Horisberger, J.~Stahn,
J. Alloys and
  Compounds \textbf{509}, 8283  (2011).


\bibitem{2015:PRB:MG:FeN}
A.~Tayal, M.~Gupta, A.~Gupta, P.~Rajput, J.~Stahn, Phys. Rev. B
\textbf{92},
  054109 (2015).


\bibitem{2015_JAC_FeAlN_MG}
A.~Tayal, M.~Gupta, N.~Pandey, A.~Gupta, M.~Horisberger, J.~Stahn,
Journal of
  Alloys and Compounds \textbf{650}, 647  (2015).


\bibitem{2017_Mat_today_FeN_dcMS}
Materials Today: Proceedings \textbf{4}, 6173  (2017).


\bibitem{2019_JMMM_Fe4N_by_FeN}
Q.~Lu, M.~Xie, G.~Han, B.~Zheng, Y.~Song, J.~Qiang, X.~Wang,
Z.~Wu, P.~Yan,
  W.~Liu, Journal of Magnetism and Magnetic Materials \textbf{474}, 76  (2019).


\bibitem{Gupta_JAC01}
M.~Gupta, A.~Gupta, P.~Bhattacharya, P.~Misra, L.~Kukreja, J.
Alloys and
  Compounds \textbf{326}, 265  (2001).


\bibitem{2005_MSE_B_Laser_Nitriding}
F.~Sun, J.~Liu, Y.~Yang, H.~Yu, Materials Science and Engineering:
B
  \textbf{122}, 29  (2005).


\bibitem{2012_HI_FeN_PLD}
R.~Usui, Y.~Yamada, Y.~Kobayashi, Hyperfine Interactions
\textbf{205}, 13
  (2012).


\bibitem{2013_HI_FeN_PLD}
Y.~Yamada, R.~Usui, Y.~Kobayashi, Hyperfine Interactions
\textbf{219}, 13
  (2013).


\bibitem{SCT15_Akhil}
A.~Tayal, M.~Gupta, A.~Gupta, V.~Ganesan, L.~Behera, S.~Singh,
S.~Basu, Surface
  and Coatings Technology \textbf{275}, 264 (2015).


\bibitem{Navio.PRB08}
C.~Nav\'{\i}o, J.~Alvarez, M.J. Capitan, F.~Yndurain, R.~Miranda,
Phys. Rev. B
  \textbf{78}, 155417 (2008).


\bibitem{2003_HI_Borsa}
D.M. Borsa, D.O. Boerma, Hyperfine Interactions \textbf{151}, 31
(2003).


\bibitem{2006_NIMB_stoichiometic_FeN}
E.~Andrzejewska, R.~Gonzalez-Arrabal, D.~Borsa, D.~Boerma, Nuclear
Instruments
  and Methods in Physics Research Section B: Beam Interactions with Materials
  and Atoms \textbf{249}, 838  (2006).


\bibitem{JPak_2010}
J.~Pak, W.~Lin, K.~Wang, A.~Chinchore, M.~Shi, D.C. Ingram, A.R.
Smith, K.~Sun,
  J.M. Lucy, A.J. Hauser, F.Y. Yang, Journal of Vacuum Science \& Technology A
  \textbf{28}, 536 (2010).


\bibitem{JCG15_FeN_MBE_Naito}
M.~Naito, K.~Uehara, R.~Takeda, Y.~Taniyasu, H.~Yamamoto, Journal
of Crystal
  Growth \textbf{415}, 36 (2015)

\bibitem{2017_Iorg_Chem_FeN_FeN2}
K.~Niwa, T.~Terabe, D.~Kato, S.~Takayama, M.~Kato, K.~Soda,
M.~Hasegawa,
  Inorganic Chemistry \textbf{56}, 6410 (2017).


\bibitem{2018_JAC_FeN_megabar}
D.~Laniel, A.~Dewaele, S.~Anzellini, N.~Guignot, Journal of Alloys
and
  Compounds \textbf{733}, 53  (2018).

\bibitem{2017_Angew_Chem_NiAs_FeN}
W.P. Clark, S.~Steinberg, R.~Dronskowski, C.~McCammon, I.~Kupenko,
M.~Bykov,
  L.~Dubrovinsky, L.G. Akselrud, U.~Schwarz, R.~Niewa, Angewandte Chemie
  International Edition \textbf{56}, 7302 (2017).


\bibitem{2006_Cat_Comm_FeN_Catalysis}
M.~Zheng, X.~Chen, R.~Cheng, N.~Li, J.~Sun, X.~Wang, T.~Zhang,
Catalysis
  Communications \textbf{7}, 187  (2006).


\bibitem{2011_EES_FeN_Catalyst}
Z.~Yang, S.~Guo, X.~Pan, J.~Wang, X.~Bao, Energy Environ. Sci.
\textbf{4}, 4500
  (2011).


\bibitem{Navio:APL:2009}
C.~Nav\'{\i}o, J.~Alvarez, M.J. Capitan, J.~Camarero, R.~Miranda,
Appl. Phys.
  Lett. \textbf{94}, 263112 (2009).


\bibitem{Navio.NJP2010}
C.~Nav\'{\i}o, M.J. Capit\'{a}n, J.~\'{A}lvarez, R.~Miranda,
F.~Yndurain, New
  J. Phys. \textbf{12}, 073004 (2010)

\bibitem{2019_JAC_FeN_1st_pricipal}
S.~Li, X.~Cui, Z.F. Liu, T.L. Song, Journal of Alloys and
Compounds
  \textbf{771}, 322  (2019).
\newblock

\bibitem{2017_JPCS_FeN_CoN_DFT}
Journal of Physics and Chemistry of Solids \textbf{100}, 148
(2017).


\bibitem{2015_RSC_ADV_mag_order}
Z.~Zhao, K.~Bao, D.~Duan, F.~Tian, B.~Liu, T.~Cui, RSC Adv.
\textbf{5}, 31270
  (2015).


\bibitem{2014_JAC_3d_mononitride_1st_princ}
Journal of Alloys and Compounds \textbf{603}, 172  (2014).


\bibitem{2012_JAC_Soni}
Journal of Alloys and Compounds \textbf{522}, 106  (2012).


\bibitem{FeN:PRB:Houari}
A.~Houari, S.F. Matar, M.A. Belkhir, M.~Nakhl, Phys. Rev. B
\textbf{75}, 064420
  (2007).


\bibitem{1997_SSR_JG_Chen_Review}
J.G. Chen, Surface Science Reports \textbf{30}, 1  (1997).


\bibitem{1996_HI_Ruffer_NRS}
R.~R{\"u}ffer, A.I. Chumakov, Hyperfine Interactions \textbf{97},
589
  (1996).


\bibitem{XAS_beamline}
D.M. Phase, M.~Gupta, S.~Potdar, L.~Behera, R.~Sah, A.~Gupta, AIP
Conference
  Proceedings \textbf{1591}, 685 (2014).


\bibitem{p01c}
Dynamics beamline p01 web site.
\newblock
  \url{http://photon-science.desy.de/facilities/petra_iii/beamlines/p01_dynamics/index_eng.html}

\bibitem{2010_JAC_Cr_XAS_insitu}
Journal of Alloys and Compounds \textbf{508}, 233  (2010).


\bibitem{2014_BJN_TiO2_insitu_XAS}
M.~Tallarida, C.~Das, D.~Schmeisser, Beilstein J. Nanotechnol.
\textbf{5}, 77
  (2014).


\bibitem{2017_AIP_TiN_RG}
R.~Gupta, M.~Gupta, D.M. Phase, AIP Conference Proceedings
\textbf{1832},
  080081 (2017).


\bibitem{2019_JPCM_Fe_insitu}
P.~Vishwakarma, M.~Gupta, D.M. Phase, A.~Gupts, Journal of
Physics: Condensed
  Matter \textbf{31}, 105001 (2019).


\bibitem{wille2010NFS}
H.~Wille, H.~Franz, R.~R{\"o}hlsberger, W.~Caliebe, F.~Dill, in
Journal
  of Physics: Conference Series \textbf{217} 012008 (2010).

\bibitem{2013_Kai_NRS}
Journal of Electron Spectroscopy and Related Phenomena
\textbf{189}, 187
  (2013).


\bibitem{1996_HI_Smirnov_NRS_SR}
G.V. Smirnov, Hyperfine Interactions \textbf{97}(1), 551 (1996).


\bibitem{2008_CRP_NRS_Ruffer}
Comptes Rendus Physique \textbf{9}(5), 595  (2008).


\bibitem{Ralf_NRS}
R.~R\"{o}hlsberger, \emph{Nuclear Condensed Matter Physics with
Synchrotron
  Radiation}, vol. 208 Springer-Verlag (2004).

\bibitem{Naito:FeN:14}
Journal of Crystal Growth \textbf{415}, 36  (2015).


\bibitem{1993_JESRP_SORIANO_TiN_XAS}
Journal of Electron Spectroscopy and Related Phenomena
\textbf{62}, 197
  (1993).


\bibitem{2004_SSC_N_K_edge_CrN_C2N}
Solid State Communications \textbf{130}(3), 209  (2004).


\bibitem{2019_TSF_CrN_MG}
Y.~Tripathi, R.~Gupta, Seema, M.~Gupta, D.~Phase, P.~Rajput, Thin
Solid Films
  \textbf{670}, 113  (2019).


\bibitem{2018_JVST_NiO_LT}
Y.~Abe, S.~Yamauchi, M.~Kawamura, K.H. Kim, T.~Kiba, Journal of
Vacuum Science
  \& Technology A \textbf{36}, 02C102 (2018).


\bibitem{PRB_DB_1}
U.~van B{\"u}rck, W.~Potzel, P.~Schindelmann, Y.V. Shvyd'ko,
E.~Gerdau,
  O.~Leupold, H.~R{\"u}ter, Physical Review A \textbf{61}, 013803
  (1999).

\bibitem{PRB_DB_2}
Y.V. Shvyd'ko, U.~Van~B{\"u}rck, W.~Potzel, P.~Schindelmann,
E.~Gerdau,
  O.~Leupold, J.~Metge, H.~R{\"u}ter, G.~Smirnov, Physical Review B
  \textbf{57}(6), 3552 (1998).

\bibitem{PRL_QB_1}
E.~Gerdau, R.~R{\"u}ffer, R.~Hollatz, J.~Hannon, Physical review
letters
  \textbf{57}(9), 1141 (1986).

\bibitem{REFTIM2008}
M.A. Andreeva, Hyperfine Interactions \textbf{185}(1), 17 (2008).


\end{thebibliography}

\end{document}